\documentstyle[12pt]{article}

\newcounter{rozd}
\newcounter{rown}[rozd]

\newtheorem{df}{Definition}[rozd]

\newtheorem{prop}[df]{Proposition}
\newtheorem{lemma}[df]{Lemma}
\newtheorem{tw}[df]{Theorem}
\newtheorem{cor}[df]{Corollary}

\newtheorem{rem}[df]{Remark}

\newcommand{\rozdzial}{\stepcounter{rozd}
                       \stepcounter{rown}
                       \section}
\newcommand{\nr}{\eqno(\therozd.\therown)\stepcounter{rown}}

\def\bbbc{{\mathchoice {\setbox0=\hbox{$\displaystyle\rm C$}
   \hbox{\hbox to0pt{\kern0.4\wd0\vrule height0.9\ht0\hss}
   \box0}}{\setbox0=\hbox{$\textstyle\rm C$}\hbox{\hbox
   to0pt{\kern0.4\wd0\vrule height0.9\ht0\hss}\box0}}
   {\setbox0=\hbox{$\scriptstyle\rm C$}\hbox{\hbox
   to0pt{\kern0.4\wd0\vrule height0.9\ht0\hss}\box0}}
   {\setbox0=\hbox{$\scriptscriptstyle\rm C$}\hbox{\hbox
   to0pt{\kern0.4\wd0\vrule height0.9\ht0\hss}\box0}}}}
\def\bbbn{{\rm I\!N}}
\def\squareforqed{\hbox{\rlap{$\sqcap$}$\sqcup$}}
\def\bbbz{{\mathchoice {\hbox{$\sf\textstyle Z\kern-0.4em Z$}}
   {\hbox{$\sf\textstyle Z\kern-0.4em Z$}}
   {\hbox{$\sf\scriptstyle Z\kern-0.3em Z$}}
   {\hbox{$\sf\scriptscriptstyle Z\kern-0.2em Z$}}}}
\def\bbbone{{\mathchoice {\rm 1\mskip-4mu l} {\rm 1\mskip-4mu l}
   {\rm 1\mskip-4.5mu l} {\rm 1\mskip-5mu l}}}
\def\squareforqed{\hbox{\rlap{$\sqcap$}$\sqcup$}}

\newcommand{\cA}{{\cal A}}
\newcommand{\cB}{{\cal B}}

\newcommand{\cF}{{\cal F}}
\newcommand{\cH}{{\cal H}}

\newcommand{\cK}{{\cal K}}
\newcommand{\cL}{{\cal L}}

\newcommand{\cP}{{\cal P}}
\newcommand{\cPc}{{\cal P}_{\rm C}}
\newcommand{\cPd}{{\cal P}_{\rm D}}
\newcommand{\cS}{{\cal S}}
\newcommand{\cSs}{{\cal S}_{\rm sep}}
\newcommand{\cSsd}{{\cal S}_{\rm sep}^{\rm d}}
\newcommand{\cSt}{{\cal S}_\tau}
\newcommand{\cStd}{{\cal S}_\tau^{\rm d}}
\newcommand{\cT}{{\cal T}}

\newcommand{\BH}{{\cB(\cH)}}
\newcommand{\BK}{{\cB(\cK)}}
\newcommand{\vr}{\varrho}

\newcommand{\id}{{\rm id}}
\newcommand{\jed}{\bbbone}

\newcommand{\tens}{\otimes}
\newcommand{\Tens}{{\cB(\cH)\tens\cB(\cK)}}
\newcommand{\Tensh}{{\cH\tens\cK}}

\newcommand{\Tr}{{\rm Tr}\,}

\newcommand{\beag}{\begin{eqnarray*}}
\newcommand{\eeag}{\end{eqnarray*}}
\newcommand{\bea}{\begin{eqnarray}}
\newcommand{\eea}{\end{eqnarray}}
\newcommand{\ba}{\begin{array}}
\newcommand{\ea}{\end{array}}
\renewcommand{\[}{\stepcounter{rown}\begin{equation}}
\renewcommand{\]}[1]{\label{#1}\end{equation}}

\newcommand{\proof}{\par\noindent {\it Proof. }}
\newcommand{\qed}{\hfill\squareforqed\vspace{2mm}\par\noindent}

\newcommand{\Jm}{J_{\mathrm{m}}}

\newcommand{\rr}{\vr^{1/2}}

\title{On a characterization of PPT states\thanks{Supported
by the grant SCALA (IST-2004-015714) 
}}
\author{W{\l}adys{\l}aw A. Majewski \\
Institute of Theoretical Physics and Astrophysics \\
Gda\'nsk University \\ Wita Stwosza 57 \\ 80-952 Gda\'nsk, Poland \\
{\normalsize {\it e-mail:} {\tt fizwam@univ.gda.pl}}
}
\begin{document}
\maketitle

\begin{abstract}
We present two different descriptions of positive partially transposed (PPT) states.  One is based on the theory of positive maps while the second description provides a characterization of PPT states in terms of Hilbert space vectors. 
Our note is based on our previous results presented in \cite{I}, \cite{II}, and \cite{MajOSID}.
\end{abstract}

\rozdzial{Definitions and notations}
In Quantum Computing a characterization of states with positive partial transposition is important problem (see \cite{Hor}).  Recently, some partial results in this direction were obtained (see \cite{CK1}, and \cite{CK2}). The aim of this note, based on our previous results
(see \cite{I}, \cite{II}, and \cite{MajOSID}), is to present two different {\it complete} characterization of PPT states.

For the sake of convenience, we provide all necessary preliminaries and set up the notation.
Let $\cB(\cH)$ be the set of all linear bounded operators on a Hilbert space $\cH$.
We denote the set of all positive
elements of $\cB(\cH)$ by $\cB(\cH)^+$. A \textit{state}
on $\cB(\cH)$ is a linear functional $\phi:\cB(\cH)\longrightarrow\bbbc$
such that $\phi(A)\geq 0$ for every $A\in\cB(\cH)^+$ and $\phi(\jed)=1$,
where $\jed$ is the unit of $\cB(\cH)$.
The set of all states on $\cB(\cH)$ is denoted by $\cS_{\cB(\cH)}$. For any subset
$\cT\subset\cS_\cH$ we define the \textit{dual cone} by
$$
\cT^{\rm d}=\{A\in\cB(\cH):\mbox{$\phi(A)\geq 0$ for every $\phi\in\cT$}\}.
$$
It is easy to check that the definition of a state implies
$\cB(\cH)^+\subset\cT^{\rm d}$ for every $\cT\subset\cS_{\cB(\cH)}$. We say that
the family $\cT$ determines the order of $\cB(\cH)$ when $\cT^{\rm d}=\cB(\cH)^+$.

A linear map $\Psi:\cB(\cH_1)\longrightarrow\cB(\cH_2)$ is called \textit{positive} if
$\Psi(\cB(\cH_1)^+)\subset\cB(\cH_2)^+$.
For $k\in\bbbn$ we consider a map
$\Psi_k:M_k\tens\cB(\cH_1)\longrightarrow M_k\tens\cB(\cH_2)$ where
$M_k$ denotes the algebra of $k\times k$-matrices with complex entries
and $\Psi_k= \id_{M_k}\tens\Psi$. We say that $\Psi$ is
\textit{$k$-positive} if the map $\Psi_k$ is positive.
The map $\Psi$ is said \textit{completely positive} when $\Psi$ is
$k$-positive for every $k\in\bbbn$.
Let us recall that for a finite dimensional Hilbert space $\cL$
every state $\phi$ on $\cB(\cL)$ has the form of
$\phi(A)=\Tr(\vr A)$, where $\vr$ is a uniquely determined
\textit{density matrix},
i.e. an element of $\cB(\cL)^+$ such that $\Tr\vr=1$.

Throughout this note $\cH$ and $\cK$ will be fixed finite-dimensional
Hilbert spaces.
We also fix orthonormal bases $\{e_i\}_{i=1}^n$ and $\{f_j\}_{j=1}^m$
of the spaces $\cH$ and $\cK$ respectively, where $n=\dim\cH$,
$m=\dim\cK$. For simplicity we will write $\cS$, $\cS_\cH$, $\cS_\cK$
instead of
$\cS_\Tens$, $\cS_\BH$, $\cS_\BK$, respectively. By $\tau_\cH$, $\tau_\cK$,
$\tau_\Tensh$
we denote transposition maps on $\BH$, $\BK$, $\cB(\Tensh)$, respectively,
associated with bases $\{e_i\}$, $\{f_j\}$, $\{e_i\tens f_j\}$,
respectively.
Let us note that for every finite dimensional Hilbert space $\cL$ the
transposition $\tau_\cL:\cB(\cL)\longrightarrow\cB(\cL)$ is a positive
map but not completely positive (in fact it is not even $2$-positive).

A positive map $\Psi:\BH\longrightarrow\BK$ is called
\textit{decomposable}
if
there are completely positive maps $\Psi_1,\Psi_2:\BH\longrightarrow\BK$
such that $\Psi=\Psi_1+\Psi_2\circ\tau_\cH$. Let $\cP$, $\cPc$ and $\cPd$
denote the set of all positive, completely positive
and decomposable maps from $\BH$ to $\BK$, respectively. Note that
$$ %\be
\cPc \subset \cPd \subset \cP
\nr
$$ %\ee{inclP}
(see also \cite{Ch1}).

A state $\varphi\in\cS$
is said to be \textit{separable} if it can be written in the
form
$$ %\be
\varphi=\sum_{n=1}^Na_n\varphi_n^\cH\tens\varphi_n^\cK
$$ %\ee{sep}
where $N\in\bbbn$, $\varphi_n^\cH\in\cS_\cH$, $\varphi_n^\cK\in\cS_\cK$
for $n=1,2,\ldots,N$, $a_n$ are positive numbers such that
$\sum_{n=1}^Na_n=1$, and the state $\varphi_n^\cH\tens\varphi_n^\cK$
is defined as $\varphi_n^\cH\tens\varphi_n^\cK(A\tens B)=
\varphi_n^\cH(A)\varphi_n^\cK(B)$ for $A\in\BH$, $B\in\BK$. The set of
all separable states on the algebra
$\Tens$ is denoted by $\cSs$. A state which is not
in $\cSs$ is called \textit{entangled} or \textit{non-separable}.

Finally, let us define the family of \textit{PPT} (transposable) states on
$\cB(\Tensh)$
$$ %\be
\cSt=\{\varphi\in\cS:\mbox{$\varphi\circ(\id_\BH\tens\tau_\cK)\in\cS$}\}.
$$ %\ee{St}
Note that due to the positivity of the transposition $\tau_\cK$ every
separable state $\varphi$ is transposable, so
$$ %\be
\cSs\subset\cSt\subset\cS.\nr
$$ %\ee{inclS}

As it was mentioned, in this note we provide two characterization of PPT states. The first one is based on the theory of decomposable maps. This will be done in Section 2. The Section 3 gives a quick review of Tomita-Takesaki theory for finite dimensional case and establishes  relations among transposition, modular operator and modular conjugation.
The second characterization characterization of PPT states is presented in Section 4. It follows from Tomita-Takesaki theory and a correspondence 
between density matrices (normal states) and vectors in the natural cone.
 
\rozdzial{Decomposable maps and their relation to PPT states}
 In the sequel we assume
that both finite dimensional Hilbert spaces $\cH$ and $\cK$ have dimension greater than $1$.

For any element $x\in\cH$, define the linear operator
$V_x:\cK\longrightarrow
\Tensh$ by $V_xz=x\tens z$ for $z\in \cK$. By $E_{x,y}$ where
$x,y\in\cH$
we denote
the one-dimensional operator on $\cH$ defined by
$E_{x,y}u=\langle y,u\rangle x$ for $u\in\cH$.
For simplicity, if $\{e_i\}_{i=1}^n$ is a basis of $\cH$, we write
$V_i$ and $E_{ij}$ instead of $V_{e_i}$ and $E_{e_i,e_j}$ for any
$i,j=1,2,\ldots,n$.

It is not hard (see \cite{I}) to show the following equality
$$
H=\sum_{i,j=1}^nE_{ij}\tens V_i^*HV_j.
$$
which is nothing but a form of decomposition of $H \in \cB(\cH \otimes \cK)$.
This suggests that for a fixed $H$ one can define the
map $S_H:\BH\longrightarrow\BK$
\begin{equation}
\label{2.1}
S_H(E_{x,y})=V_x^*HV_y
\end{equation}
where $x,y\in\cH$. The correspondence between $H$ and $S_H$ was observed
by Choi (see \cite{Ch3}).

As the first step, we wish to describe the properties of positive decomposable
maps. It will be done by means of the family of PPT states.
We will need the following lemma proved in \cite{I}:
\begin{lemma}
\label{fi}
Let $k\in\bbbn$ and $A\in M_k\tens\BH$. Suppose that both $A$ and
$(\tau_{M_k}\tens\id_\BH)(A)$ are positive in $M_k\tens\BH$. Then for every
vectors $x_1,x_2,\ldots, x_k\in\cK$ the map
$\psi:\cB(\Tensh)\longrightarrow\bbbc$ defined as
$$
\psi(C)=\sum_{i,j=1}^k\sum_{p,r=1}^n\langle h_i\tens e_p, Ah_j
\tens e_r\rangle\langle e_p\tens x_i, Ce_r\tens x_j\rangle,\;\;\;\;\;
C\in\cB(\Tensh)
$$
is a positive functional on $\cB(\Tensh)$ such that
$\psi\circ(\tau_\cH\tens\id_\BK)$ is also positive.
\end{lemma}

For the reader's convenience, here, we reproduce the proof given in \cite{I}.
\proof
First of all note that
for every state $\varphi\in\cS$ we have by definition:
$$
\varphi\in\cSt\Longleftrightarrow\varphi\circ
(\tau_\cH\tens\id_\BK)\in\cS.
$$
Observe that
\beag
\psi(C)
&=& \sum_{i,j,p,r}\langle h_i\tens e_p\tens e_p\tens x_i,
    (A\tens C) h_j\tens e_r\tens e_r\tens x_j\rangle \\
&=& \left\langle\sum_{i,p}h_i\tens e_p\tens e_p\tens x_i,(A\tens C)
    \sum_{i,p}h_i\tens e_p\tens e_p\tens x_i\right\rangle.
\eeag
If $C$ is positive then $A\tens C$ is positive in the algebra
$M_k\tens\BH\tens\BH\tens\BK$, so $\psi(C)\geq 0$.
On the other hand,
\beag
\psi(\tau_\cH\tens\id_\BK)(C)
&=& \sum_{i,j,p,r}\langle h_i\tens e_p,Ah_j\tens e_r\rangle\langle
    e_r\tens x_i,Ce_p\tens x_j\rangle \\
&=& \sum_{i,j,p,r}\langle h_i\tens e_r,(\id_{M_k}\tens\tau_\cH)(A)
    h_j\tens e_p\rangle\langle e_r\tens x_i,Ce_p\tens x_j\rangle \\
&=& \left\langle\sum_{i,r}h_i\tens e_r\tens e_r\tens x_i,[(\id_{M_k}
    \tens\tau_\cH)(A)\tens C]\sum_{i,r}h_i\tens e_r\tens e_r\tens x_i
    \right\rangle.
\eeag
The positivity of $(\tau_{M_k}\tens\id_\BH)(A)$
implies the positivity of $(\id_{M_k}\tens\tau_\cH)(A)$, so by the
above arguments, if $C$ is positive then
$\psi(\tau_\cH\tens\id_\BK)(C)\geq 0$.
\qed

The main result, taken from \cite{I}, is
\begin{tw}
\label{Hrho}
For any selfadjoint operator $H$ the map $S_H$ is decomposable if and only if
$H\in\cStd$.
\end{tw}
Again, for the reader's convenience, we reproduce the proof given in \cite{I}.

\proof
Suppose that
$S_H=S_1+S_2\circ\tau_\cH$, where $S_1$, $S_2$ are completely positive.
Then $H=H_1+(\tau_\cH\tens\id_\BK)(H_2)$ where $H_1$, $H_2$ are positive
operators such that $S_i=S_{H_i}$, $i=1,2$. Let $\varphi\in\cSt$. Hence,
$$
\varphi(H)=\varphi(H_1)+\varphi(\tau_\cH\tens\id_\BK)(H_2)\geq 0
$$
because both $\varphi$ and $\varphi(\tau_\cH\tens\id_\BK)$ are positive
functionals.

Conversely, let $H\in\cStd$. Suppose that $K\in\bbbn$ and
$A=[A_{ij}]_{i,j=1,2,\ldots,k}\in M_k\tens\BH$ is such that both
$A$ and $(\tau_{M_k}\tens\id_\BH)(A)$ are positive in $M_k\tens\BH$.
>From the theorem of St{\o}rmer (\cite{St3}, see also \cite{Rob}) it is
enough to
show that $(\id_{M_k}\tens S_H)(A)$ is a positive element in
$M_k\tens\BK\simeq\cB(\bbbc^k\tens\cK)$. To this end let us fix an element
$h\in\bbbc^k\tens\cK$. Let $h=\sum_{s=1}^kh_s\tens x_s$. Then
\beag
\langle h,(\id_{M_k}\tens S_H)(A)h\rangle
&=& \sum_{s,t}\sum_{i,j}\sum_{p,r}\langle e_p, A_{ij}e_r\rangle
    \langle h_s\tens x_s, (F_{ij}\tens V_p^*HV_r)h_t\tens x_t\rangle \\
&=& \sum_{s,t}\sum_{i,j}\sum_{p,r}\langle e_p, A_{ij}e_r\rangle\langle
    h_s, F_{ij}h_t\rangle\langle e_p\tens x_s, He_r\tens x_t\rangle \\
&=& \sum_{i,j}\sum_{p,r}\langle e_p,A_{ij}e_r\rangle\langle
    e_p\tens x_i,He_r\tens x_j\rangle
\eeag
where $F_{ij}$'s are matrix units in $M_k$.
The last expression is nonnegative by Lemma \ref{fi}.
\qed

The above theorem leads to the first promised characterization of PPT states (see also \cite{Ch1}, \cite{J}, \cite{MROMP}, \cite{I}). It is worth pointing out that a similar characterization was obtained in \cite{L1} and \cite{L2}.
\begin{cor}
\begin{enumerate}
\item
Let $\cP_D$ be the set of all decomposable maps. Denote by $\cF_H$ the corresponding set of their self-adjoint operators (given by the Choi's correspondence, see (\ref{2.1})). Then $(\cF_H)^d$ is the set of all PPT states.
\item
Let $\cP$ be the set of all positive maps. Denote by $\cF_H^0$ the corresponding set of their self-adjoint operators (given by the Choi's correspondence, see (\ref{2.1})). Then $(\cF_H^0)^d$ is the set of all separable states.
\end{enumerate}
\end{cor}
 
 Having such characterization the following remarks are relevant:
\begin{rem}
\begin{enumerate}
\item
In \cite{Ha,Hor,KK,Rob, Wor} it was shown that, in general,
$\cStd$ is a proper subset of
$\cSsd$.
\item
Although the above corollary provides a natural characterization of PPT states one could ask for more ``effective'' characterization of the considered families of states. This question can be taken as a motivation for another characterization which will be presented in Section 4.
\end{enumerate}
\end{rem}

We end this section with one observation (for others see \cite{I}) shedding some new light on the complicated structure of PPT states.
To quote this result,
whose principal significance is that it allows
one to write (or verify) concrete examples of PPT states,
we need to recall the following
\begin{lemma} (\cite{Ch2,Rob})
\label{uw}
Suppose that $\vr=[\vr_{ij}]$ is an operator on $\Tensh$. One has:
$[\vr_{ij}]$ is positive if and only if the matrix
$[\tilde{\vr_{ij}}]_{i,j=1,2,\ldots,n-1}$ is positive, where
$\tilde{\vr_{ij}}=\vr_{ij}-\vr_{in}\vr_{nn}^{-1}\vr_{nj}$
for $i,j=1,2,\ldots,n-1$. 
\end{lemma}
Therefore, replacing $\vr_{nn}$ by $\vr_{nn}
+\varepsilon\jed$ if necessary we may suppose that $\vr_{nn}$ is invertible
and then by an application of the above lemma we can restrict ourselves
to the case of two-dimensional space $\cH$. Thus

\begin{prop}(\cite{I})
Let $\dim\cH=2$ and let $\vr=[\vr_{ij}]_{i,j=1,2}$
($\vr_{ij}\in\BK$ as in the above lemma) be a density
matrix on the space
$\Tensh$. Assume that there exists a vector $f\in \cH$ and a selfadjoint
operator
$A$ on $\cH$ with the property
$$\langle f\tens y,\{A\tens \jed,\vr\} f\tens y\rangle=0$$
for any $y\in\cK$, where $\{\cdot,\cdot\}$ stands for the anticommutator.
Then $\vr^\tau=[\vr_{ji}]_{i,j=1,2}$ is also
positive.
\end{prop}

\rozdzial{Tomita-Takesaki scheme for transposition}\label{Tomita}
In this section, reproduced from \cite{II}, we indicate how Tomita-Takesaki techniques may be used to describe a transposition. Although, from the very mathematical point of view one can consider this Section as an exercise on Tomita-Takesaki theory, the presented results clearly indicate how transposition is related to modular conjugation and modular operator. Moreover, 
it gives a sufficient preparation for the next section where the second characterization of PPT states will be given. For a comprehensive account of Tomita-Takesaki theory addressed to physicists we refer Haag's book \cite{Haag}, while mathematical description can be found in \cite{Araki}, and \cite{Takesaki}.

Let $\cH$ be a finite dimensional (say $n$-dimensional) Hilbert space. 
Define $\omega\in \cS_{\cB(\cH)}$ as
$
\omega(a)= \Tr\varrho a,
$
where $\varrho$ is an invertible density matrix, i.e. the state $\omega$
is a faithful one.
Denote by $(\cH_\pi,\pi,\Omega)$ the GNS triple associated with
$(\cB(\cH),\omega)$.
Then, one has:
\begin{itemize}
\item
$\cH_\pi$ is identified with $\cB(\cH)$ where the inner product
$(\cdot\,,\cdot)$ is defined
as $(a,b)=\Tr a^*b$, $a,b\in \cB(\cH)$;
\item
With the above identification: $\Omega= \rr$;
\item
$\pi(a)\Omega=a\Omega$;
\item
The modular conjugation $\Jm$ is the hermitian involution:
$\Jm a \rr =\rr a^*$;
\item
The modular operator $\Delta$ is equal to the map
$\varrho \cdot\varrho^{-1}$;
\end{itemize}

As a next step let us define two conjugations: $J_c$ on
$\cH$ and $J$ on $\cH_\pi$.
To this end we note that the eigenvectors $\{x_i\}$ of $\varrho
=\sum_i \lambda_i |x_i\rangle \langle x_i|$ form an orthonormal basis in $\cH$
(due to the faithfulness of $\omega$). Hence we can define
\begin{equation}\label{Jcdef}
J_c f = \sum_i \overline{\langle x_i,f\rangle} x_i
\end{equation}
for every $f\in \cH$.
Due to the fact that $E_{ij}\equiv |x_i\rangle \langle x_j|\}$ form an 
orthonormal basis in $\cH_\pi$ we can define in the similar way a conjugation $J$ on
$\cH_\pi$
\begin{equation}\label{Jdef}
J a \rr = \sum_{ij} \overline{(E_{ij},a \rr)} E_{ij}
\end{equation}
Obviously, $J\rr = \rr$. 

Now let us define a transposition on $\cB(\cH)$ as the map $a\mapsto a^t\equiv J_ca^*J_c$ where
$a\in \cB(\cH)$. By $\tau$ we will denote the map induced on $\cH_\pi$ by the transposition, i.e.
\begin{equation}\label{tau}
\tau a\rr=a^t\rr
\end{equation}
where $a\in \cB(\cH)$.
The main properties of the notions introduced above are the following
\begin{prop} (\cite{II})
\label{transposition}
Let $a\in \cB(\cH)$ and $\xi\in \cH_\pi$. Then
$$a^t\xi=Ja^*J\xi.$$
\end{prop}
\begin{proof}
Let $\xi=b\rr$ for some $b\in \cB(\cH)$. Then we can perform the following cal\-cu\-la\-tions
\begin{eqnarray*}
\lefteqn{Ja^*Jb\rr=}\\
&=&
\sum_{ij}\overline{(E_{ij},a^*Jb\rr)}E_{ij}=%\\&=&
\sum_{ij}\sum_{kl}(E_{kl},b\rr)\overline{(E_{ij},a^*E_{kl})}E_{ij}\\
&=&
\sum_{ijkl}\Tr (E_{lk}b\rr)\overline{\Tr (E_{ji}a^*E_{kl})}E_{ij}=%\\&=&
\sum_{ijk}\Tr (E_{jk}b\rr)\overline{\Tr (E_{ki}a^*)}E_{ij}\\
&=&
\sum_{ijk}\langle x_k,b\rr x_j\rangle\overline{\langle x_i,a^*x_k\rangle}E_{ij}=%\\&=&
\sum_{ijk}\langle J_cb\rr x_j,x_k\rangle\langle x_k,ax_i\rangle E_{ij}\\
&=&
\sum_{ij}\langle J_cb\rr x_j,ax_i\rangle E_{ij}=
\sum_{ij}\langle a^*J_cb\rr x_j,x_i\rangle E_{ij}\\
&=&
\sum_{ij}\langle x_i,J_ca^*J_cb\rr x_j\rangle E_{ij}=
\sum_{ij}\langle x_i,a^tb\rr x_j\rangle E_{ij}\\
&=&
\sum_{ij}\Tr (E_{ji}a^tb\rr)E_{ij}=
\sum_{ij}(E_{ij},a^tb\rr)E_{ij}=
a^tb\rr
\end{eqnarray*}
\qed
\end{proof}

As a next step let us consider the modular conjugation $\Jm$ which
has the form
\begin{equation}\label{Jm}
\Jm a \rr = (a\rr)^* = \rr a^*
\end{equation}
%However, we observe
%\begin{eqnarray*}
%\rr a^* &=& \sum_{ij} (E_{ij}, \rr a^*) E_{ij}
%=\sum_{ij}(\Tr E_{ji}\rr a^*) E_{ij}=\\
%&=& \sum_{ij}(x_i, \rr a^* x_j) E_{ij} =
%\sum_{ij} \lambda_i^{1/2} (x_i,a^* x_j) E_{ij}.
%\end{eqnarray*}
Define also the unitary operator $U$ on $\cH_\pi$ by
\begin{equation}\label{U}
U = \sum_{ij} |E_{ji}\rangle \langle E_{ij}|
\end{equation}
Clearly, $UE_{ij} = E_{ji}$. We have the following
\begin{prop}(see \cite{II})
\label{commute}
Let $J$ and $\Jm$ be the conjugations introduced above and $U$ be the unitary operator
defined by (\ref{U}). Then we have:
\begin{enumerate}
\item\label{commute1} $U^2 = \jed$ and $U = U^*$
\item\label{commute2} $J=U\Jm$;
\item\label{commute3} $J$, $\Jm$ and $U$ mutually commute.
\end{enumerate}
\end{prop}
\begin{proof}
(\ref{commute1})
We calculate
$$ \sum_{ijmn}|E_{ij}\rangle \langle E_{ji}||E_{mn}\rangle \langle E_{nm}|
=\sum_{ijmn} \Tr (E_{ij} E_{mn}) |E_{ij}\rangle \langle E_{nm}|
=\sum_{ij} |E_{ij}\rangle \langle E_{ij}| = \jed
$$
The rest is evident.

(\ref{commute2}) 
Let $b\in \cB(\cH)$. Then 
\begin{eqnarray*}
U\Jm b\rr&=&U\rr b^*=\sum_{ij}(E_{ji},\rr b^*)E_{ij}\\
&=&\sum_{ij}\Tr (E_{ij}\rr b^*)E_{ij}=
\sum_{ij}\langle x_j,\rr b^*x_i\rangle E_{ij}\\
&=&\sum_{ij}\overline{\langle x_i,b\rr x_j\rangle}E_{ij}=
\sum_{ij}\overline{\Tr (E_{ji}b\rr )}E_{ij}\\
&=&\sum_{ij}\overline{(E_{ij},b\rr)}E_{ij}=
Jb\rr
\end{eqnarray*}

(\ref{commute3}) 
$J$ is an involution, so by the previous point we have $U\Jm U\Jm=\jed$.
It is equivalent to the equality $U\Jm=\Jm U$. Hence we obtain $U\Jm=J=\Jm U$ and
consequently 
$UJ=\Jm=JU$ and $\Jm J=U=J\Jm$ because both $U$ and $\Jm$ are also 
involutions.
\qed
\end{proof}

Now, we are ready to describe a polar decomposition of the map $\tau$.
\begin{tw} (\cite{II})
\label{polar}
If $\tau$ is the map introduced in (\ref{tau}), then
$$
\tau=U\Delta^{1/2}.
$$
\end{tw}
\begin{proof}
Let $a\in \cB(\cH)$. Then by Proposition \ref{transposition} and Proposition \ref{commute}(2)
we have
$$%\begin{equation}\label{t6}
\tau a\rr = a^t \rr = J a^* J \rr = J \Jm  \Delta^{1/2}a \rr
= U \Delta^{1/2} a \rr.
$$
\qed
\end{proof}
%\begin{corollary}
%For a linear map $\varphi$ on $\cB(H)$ we have
%$$8
%(\varphi\circ t)
%$$%\begin{equation}\label{t7}
%(\varphi \circ ^{\mathrm{t}})(a) \rr \equiv
%\varphi(a^{\mathrm{t}}) \rr = T_{\varphi} U \Delta^{1/2} a \rr.
%\eqno(7)
%$$%\end{equation}

Now we wish to prove some properties of $U$ which are analogous to that of the
modular conjugation $\Jm$. To this end we firstly need the following
\begin{lemma} (\cite{II})
\label{JcomD}
$J$ commutes with $\Delta$.
\end{lemma}
\begin{proof}
Let $a\in \cB(\cH)$. Then by Propositions \ref{transposition}, \ref{commute} 
and Theorem \ref{polar} we have
\begin{eqnarray*}
\Delta^{1/2}Ja\rr&=&\Delta^{1/2}JaJ\rr=
\Delta^{1/2}(a^*)^t\rr=UU\Delta^{1/2}(a^*)^t\rr\\
&=&Ua^*\rr=UJJa^*J\rr=
JUa^t\rr\\&=&JUU\Delta^{1/2}a\rr=
J\Delta^{1/2}a\rr
\end{eqnarray*}
So, $\Delta^{1/2}J=J\Delta^{1/2}$ and consequently $\Delta J=\Delta^{1/2}J\Delta^{1/2}=
J\Delta$.
\qed
\end{proof}
We will also use (cf. \cite{Araki}) 
$$ 
V_{\beta} = \mathrm{closure} \left\{
\Delta^{\beta}a \rr:\; a \geq 0, \;\beta \in
\left[0,\frac{1}{2}\right]\right\}. 
$$ 
Clearly, each $V_{\beta}$
is a pointed, generating cone in $\cH_\pi$ and 
\begin{equation}\label{duality}
V_\beta=\{\xi\in H_\pi:\,\mbox{$(\eta,\xi)\geq 0$ for all $\eta\in V_{(1/2)-\beta}$}\}
\end{equation}
Recall that $V_{1/4}$ is nothing
but {\bf the natural cone} $\cP$ associated with the pair $(\pi(\cB(\cH)),\Omega)$
(see \cite[Proposition 2.5.26(1)]{BR}).
Finally, let us define an automorphism $\alpha$ on $\cB(\cH_{\pi})$ by
\begin{equation}
\label{alfa}
\alpha(a) = U a U^* = U a U,\quad a\in \cB(\cH_{\pi}),
\end{equation}
as $U$ is self-adjoint.
Then we have
\begin{prop} (\cite{II})
\label{przestawianie}
\begin{enumerate}
\item\label{przestawianie1} $U \Delta = \Delta^{-1} U$
\item\label{przestawianie2} $\alpha$ maps $\pi(\cB(\cH))$ onto $\pi(\cB(\cH))^{\prime}$;
\item\label{przestawianie3} For every $\beta\in [0,1/2]$ the unitary $U$ maps $V_\beta$ 
onto $V_{(1/2)-\beta}$.
\end{enumerate}
\end{prop}
\begin{proof}
(\ref{przestawianie1})
By Proposition \ref{commute} and Lemma \ref{JcomD} we have
$$
U\Delta=J\Jm\Delta=J\Delta^{-1}\Jm=\Delta^{-1}J\Jm = \Delta^{-1} U.
$$

(\ref{przestawianie2})
Let $a,b\in \cB(\cH)$ and $\xi\in \cH_{\pi}$. Then Propositions \ref{transposition} and 
\ref{commute} imply
\begin{eqnarray*}
UaUb\xi&=&J\Jm a\Jm JbJJ\xi=J\Jm a\Jm (b^*)^tJ\xi=J(b^*)^t\Jm a\Jm J\xi\\
&=&J(b^*)^tJJ\Jm a\Jm J\xi=bJ\Jm a\Jm J\xi=bUaU\xi
\end{eqnarray*}
and the proof is complete.

(\ref{przestawianie3})
Let $a,b\in \cB(\cH)^+$. Then by the point (\ref{przestawianie1}) and Theorem \ref{polar} 
we have
\begin{eqnarray*}
\lefteqn{(\Delta^\beta b \rr, U \Delta^\beta a \rr)=}\\
&=& (\Delta^\beta b\rr, \Delta^{(1/2)-\beta}U\Delta^{1/2} a\rr)=%\\&=& 
(\Delta^\beta b\rr, \Delta^{(1/2)-\beta} a^t\rr)
\end{eqnarray*}
We recall that $a\mapsto a^t$ is a positive map on $\cB(\cH)$ so
by (\ref{duality}) the last expression is nonnegative.
Hence $UV_\beta\subset V_{(1/2)-\beta}$ for every $\beta\in [0,1/2]$.
As $U$ is an involution, we get $V_{(1/2)-\beta}=U^2V_{(1/2)-\beta}\subset
UV_\beta$ and the proof is complete.
\qed
\end{proof}

\begin{cor} (\cite{II})
$U \Delta^{1/2}$ 
maps $V_0$ into itself.
\end{cor}

Summarizing, this section establishes a close relationship between
the Tomita-Takesaki scheme and transposition.
Moreover, we have the following :
\begin{prop} (\cite{II})
\label{3.7}
Let $\xi\mapsto\omega_\xi$ be the homeomorphism between the natural
cone $\cP$ and the set of normal states on $\pi(\cB(\cH))$ described in
\cite[Theorem 2.5.31]{BR}, i.e. such that
$$\omega_\xi(a)=(\xi,a\xi),\quad a\in \cB(\cH).$$
For every state $\omega$ define
$\omega^\tau(a)=\omega(a^t)$ where $a\in \cB(\cH)$.
If $\xi\in\cP$ then the unique vector in $\cP$ mapped 
into the state
$\omega_\xi^\tau$ by the homeomorphism described above, 
is equal to $U\xi$
\end{prop}
\begin{proof}
Let $\xi=\Delta^{1/4}a\Omega$ for some $a\in \cB(\cH)^+$. Then we have
\begin{eqnarray*}
(U\xi, xU \xi)=
&=&(U \Delta^{\frac{1}{4}} a \Omega, x U \Delta^{\frac{1}{4}}a \Omega)\\
&=& (\Delta^{\frac{1}{4}} U \Delta^{\frac{1}{2}}a \Omega, x \Delta^{\frac{1}{4}}
U \Delta^{\frac{1}{2}} a \Omega) \\
&=& (\Delta^{\frac{1}{4}} a^t \Omega, x \Delta^{\frac{1}{4}} a^t \Omega)\\
&=&(\Delta^{\frac{1}{4}} JaJ \Omega, x \Delta^{\frac{1}{4}} J a J \Omega)\\
&=& (x^*J \Delta^{\frac{1}{4}} a \Omega, J \Delta^{\frac{1}{4}} a \Omega)\\
&=&(\Delta^{\frac{1}{4}}a \Omega, Jx^*J \Delta^{\frac{1}{4}} a \Omega)\\
&=& (\xi, Jx^*J\xi)\\
&=& (\xi,x^t\xi) = \omega_{\xi}(x^t)\\
\end{eqnarray*}
\qed
\end{proof}

\rozdzial{PPT states on the Hilbert-space level}\label{hilbert}
In this section we present the second characterization of PPT states. The crucial point of this approach stems from the deep Connes observation (\cite{Connes}, see also Proposition \ref{3.7}) that any density matrix (so a normal state) can be uniquely (!) represented by a vector (from the natural cone) in the Hilbert space. 

Let us begin with a preliminary observation concerning separable states.
We consider a composite system $A+B$ where a subsystem $i=A,B$ is described by $(\cB(\cK_i), \cP_i, \varrho_i)$
where $\cP_i$ denotes the natural cone associated with $(\cB(\cK_i), \varrho_i)$ (cf \cite{MajOSID} and the previous Section).
In \cite{MajOSID}, using Tomita-Takesaki approach, we have derived the one-to-one correspondence between the set of normalized vectors in $\cP_A \otimes \cP_B$ and the set of all separable states, where
$$\cP_A \otimes \cP_B \equiv \mathrm{closure}\{\sum_k a_k x^{(1)}_k \otimes x^{(2)}_k, a_k \geq 0, \sum_k a_k =1, x^{(i)}_k \in \cP_i \}$$
Here we wish to extend this result and to get an analogous characterization of PPT states. So, again, we will consider a composite system $A$ plus $B$. Moreover, again, to simplify the exposition we assume that the Hilbert spaces $\cK_A$ and $\cK_B$
are finite dimensional.
Suppose that the subsystem $A$ is described by a $C^*$ algebra $\cA \equiv \cB(\cK_A)$ equipped with a faithful state $\omega_A$ (so, of the form $\omega_A(a) \equiv Tr\{ \varrho_A a\}$ where $\varrho_A$ is an invertible density matrix). Similar y, let $\cB \equiv\cB(\cK_B)$ for some Hilbert space $\cK_B$, $\varrho_B$ be
an invertible density matrix in $\cB(\cK_B)$ and $\omega_B$ be a state
on $\cB$ such that $\omega_B(b)=\Tr(b\varrho_B)$ for $b\in \cB$. By
$(\cH,\pi,\Omega)$, $(\cH_A,\pi_A,\Omega_A)$ and
$(\cH_B,\pi_B,\Omega_B)$ we denote the GNS representations of
$(\cA\otimes \cB, \omega_A\otimes\omega_B)$, $(\cA,\omega_A)$ and
$(\cB,\omega_B)$ respectively. We observe that we can make the
following identifications (cf \cite{Cu}, \cite{MajOSID}):
\begin{enumerate}
\item $\cH=\cH_A\otimes \cH_B$,
\item $\pi=\pi_A\otimes\pi_B$,
\item $\Omega=\Omega_A\otimes\Omega_B$.
\end{enumerate}
With these identifications we have $\Jm=J_A\otimes J_B$
and $\Delta=\Delta_A\otimes\Delta_B$ where $\Jm$,
$J_A$, $J_B$ are modular conjugations and $\Delta$, $\Delta_A$,
$\Delta_B$ are modular operators for $(\pi(A\otimes B)'',\Omega)$,
$(\pi_A(A)'',\Omega_A)$, $(\pi_B(B)'',\omega_B)$ respectively.
Since $\Omega_A$ and $\Omega_B$ are separating vectors, we will
write $a\Omega_A$ and $b\Omega_B$ instead of $\pi_A(a)\Omega_A$
and $\pi_B(b)\Omega_B$ for $a\in A$ and $b\in B$.
Moreover, as finite dimensionality of the corresponding Hilbert spaces was assumed, we will also identify $\pi_A(A)^{''}$ with 
$\pi(A)$, etc. Furthermore, as $\cK_B$ is a finite dimensional Hilbert space, we denote it dimension by $n$. Thus $\cB(\cK_B) \equiv \cB({\bbbc}^n) \equiv M_n({\bbbc})$. As a next step, to put some emphasis on the dimensionality of the ``reference'' subsystem $B$, by $\cP_n$ we denote the natural cone for $(\pi(A)
\otimes \cB({\bbbc}^n), \omega \otimes \omega_0 )$ where $\omega_0$
is a faithful state on $ \cB({\bbbc}^n)$. 

Finally, the
partial transposition
$({\mathrm{id}}\otimes \tau)$ on $M_n^{\pi}(\mathcal{A})\equiv \pi(A \otimes B)$ induces
an operator at the Hilbert space level, but for the sake of
simplicity we will where convenient retain the notation
$({\mathrm{id}}\otimes \tau)$ for this operator.

In order to achieve the desired characterization of PPT states we
introduce the notion of the ``transposed cone''
${\cP}_{n}^{\tau} = ({\bf{I}} \otimes U)
{\cP}_n$, where $\tau$ is transposition on $M_n({\bbbc})$
while the operator $U$ was defined in the previous Section (we
have used the following identification: for the basis $\{e_i \}_i$
in ${\bbbc}^n$ consisting of eigenvectors of $\varrho_{\omega_0}$
($\omega_0(\cdot) = Tr\{ \varrho_{\omega_0} \cdot \}$, we have the
basis $\{ E_{ij} \equiv |e_i><e_j|\}_{ij}$ in the GNS Hilbert
space associated with $(\cB({\bbbc}^n), \omega_0)$ with
$U$ defined in terms of that basis). Note that in the same basis one has 
the identification $\cB({\bbbc}^n)$ with  $M_n({\bbbc})$.

Now the natural cone ${\cP}_n$ for $\pi(A \otimes
\cB({\bbbc}^n)) = M_n^{\pi}({\mathcal{A}})$ may be realised as
$${\cP}_n = \overline{{\Delta}^{1/4}_n\{[a_{ij}]\Omega_ :
[a_{ij}] \in M_n^{\pi}({\mathcal{A}})^+\}}$$ 
(see for example \cite[Proposition 2.5.26]{BR}). We observe:
\begin{eqnarray*}
\lefteqn{
\{({\bf I}\otimes U){\Delta}^{1/4} [a_{ij}]\Omega :\; [a_{ij}] \in
M_n^{\pi}(\mathcal{A})^+\}
}\\
&=& \{ ({\Delta}^{1/4}_A \otimes U {\Delta}^{1/4}_B) \circ \sum \pi_A(a_{ij})\otimes \pi_B(E_{ij}):\,
    [a_{ij}] \in M_n^{\pi}({\mathcal{A}})^+\}\\
&=& \{({\Delta}^{1/4}_A \otimes {\Delta}^{1/4}_B U {\Delta}^{1/2}_B) \circ \sum \pi_A(a_{ij})
    \otimes \pi_B(E_{ij}):\, [a_{ij}] \in M_n^{\pi}({\mathcal{A}})^+\}\\
&=& \{{\Delta}^{1/4}[a_{ji}]\Omega :\, [a_{ij}] \in M_n^{\pi}(\mathcal{A})^+\}.
\end{eqnarray*}
Thus
$${\cP}_n^{\tau}
= \overline{{\Delta}^{1/4}\{[a_{ji}]\Omega :
[a_{ij}] \in M_n^{\pi}(\mathcal{A})^+\}}.$$
 Postponing the task of describing the
transposed cone  more adequately at the end of this
section we wish in next lemmas to employ the geometry of cones ${\cP}_n$, ${\cP}_A$, ${\cP}_B$ in a similar way as it was done in \cite{wam} and \cite{MajOSID}. Thus,
following these lines one has

\begin{lemma}(see \cite{II})
For each $n$, ${\cP}_n \cap {\cP}_n^{\tau}$ and
$\overline{co}({\cP}_n \cup {\cP}_n^{\tau})$ are dual
cones.
\end{lemma}
\begin{proof}(see \cite{II})
For any $X\subset \cH$ we denote
$X^{\mathrm{d}}=\{\xi\in \cH:\,\mbox{$(\xi,\eta)\geq 0$ for any $\eta\in X$}\}$.
To prove the lemma it is enough to observe that
${\mathcal{P}}_n^{\mathrm{d}}={\mathcal{P}}_n$ and
$({\mathcal{P}}_n^{\tau})^{\mathrm{d}}={\mathcal{P}}_n^{\tau}$.
\qed
\end{proof}

\begin{lemma} (see \cite{II})
\label{aa}
Let $n$ be given. For any $[a_{ij}] \in M_n^{\pi}({\mathcal{A}})^+$,
$\Delta^{1/4}[a_{ij}]\Omega \in {\cP}_n \cap
{\cP}_n^{\tau}$ implies $[a_{ji}] \in M_n^{\pi}(\mathcal{A})^+$.
\end{lemma}
\begin{proof} (see \cite{II})
Let $[a_{ij}] \in M_n^{\pi}(\mathcal{A})^+$ be given and assume that
$\Delta^{1/4}[a_{ij}]\Omega \in {\cP}_n \cap
{\cP}_n^{\tau}$. We observe
$$\Delta^{1/4}[a_{ji}]\Omega = ({\bf I} \otimes
U)\Delta^{1/4}[a_{ij}]\Omega \in ({\bf I} \otimes
U)({\cP}_n \cap {\cP}_n^{\tau}) = {\cP}_n
\cap {\cP}_n^{\tau} \subset {\cP}_n.$$ 
But then the
self-duality of ${\cP}_n$ alongside (\cite{BR}; 2.5.26) will
ensure that $$0 \leq (\Delta^{1/4}[a_{ji}]\Omega_n,
\Delta^{-1/4}[b_{ij}]\Omega) = ([a_{ji}]\Omega,
[b_{ij}]\Omega)$$ for each $[b_{ij}] \in (M_n^{\pi}(\mathcal{A})')^+$.
We may now conclude from (\cite{Di}; 2.5.1 or \cite{BR}; 2.3.19)
that $[a_{ji}] \geq 0$, as required.
\qed
\end{proof}
%\textbf{Is the following true?}

\begin{tw} (see \cite{II})
\label{wniosek1}
In the finite dimensional case $\{\Delta^{1/4}[a_{ij}]\Omega :
[a_{ij}] \geq 0, [a_{ji}] \geq 0\} = {\cP}_n \cap
{\cP}_n^{\tau}$.
\end{tw}
\begin{proof} (see \cite{II})
First note that in this case $\{\Delta^{1/4}[a_{ij}]\Omega :
[a_{ij}] \geq 0\} = {\cP}_n$ (cf. \cite[Proposition 2.5.26]{BR}).
Now apply the previous lemma.
\qed
\end{proof}

Thus we got:

\begin{cor}
\label{char2}
\begin{enumerate}
\item
There is one-to-one correspondence between the set of PPT states and $\cP_n \cap \cP^{\tau}_n$.
\item
There is one-to-one correspondence between the set of separable states and $\cP_A \otimes \cP_B$.
\end{enumerate}
\end{cor}
\begin{proof}
Simple application of Theorem \ref{wniosek1} and Connes' characterization of normal states (see \cite{BR}, Theorem 2.5.31).
See also Proposition \ref{3.7}. The second statement follows from \cite{MajOSID}, see also the first paragraph of this Section.
\qed
\end{proof}

\begin{rem}
\begin{enumerate}
\item As $U$ is nontrivial the above inclusion should be, in general, the proper one.
\item As $\cP_n^{\tau}$ and $\cP_n$ contains $\cP_A \otimes \cP_B$, PPT states which are not separable are characterized by vectors in $\cP_n \cap \cP_n^{\tau} \setminus \cP_A \otimes \cP_B$. Thus, Corollary \ref{char2} gives a quite effective recipe for a construction of PPT state which is not a separable one.
\item Similary, non-PPT states are characterized by vectors $\cP_n\setminus\cP_n \cap \cP^{\tau}_n$. Again, this gives a recipe for a construction of non-PPT states.
\end{enumerate}
\end{rem}
We want to close this section, as it was announced, with a more complete characterisation of the cone
${\mathcal{P}}_n\cap{\mathcal{P}}_n^\tau$. To this end we adopt the described framework for a composite system and recall that
the natural cone $\mathcal{P}$ for
$(\pi(A\otimes B)'',\Omega)$ can be defined (see \cite{BR} or \cite{Araki}) as the closure
of the set
$$
\left\{\left(\sum_{k=1}^na_k\otimes b_k\right)j_{\mathrm{m}}\left(\sum_{l=1}^na_l\otimes b_l
\right)\Omega:\,n\in{\bf N},\,a_1,\ldots,a_n\in A,\,b_1,\ldots,b_n\in B\right\}$$
where $j_{\mathrm{m}}(\cdot)=\Jm\cdot \Jm$ is the modular
morphism on $\pi(A\otimes B)''=\pi_A(A)''\otimes\pi_B(B)''$, etc.

As it was presented (see Section 3) $\cH_B$ is the closure of the set
$\{b\rr:\,b\in B\}$ and $\Omega_B$ can be identified with $\rr$.
Let $U_B$ be the unitary
operator on $\cH_B$ described in Section 3. Then we have
\begin{lemma} (see \cite{II})
$({\bf I}\otimes U_B)\mathcal{P}$ is the closure of the set
$$
\left\{\left(\sum_{k=1}^na_k\otimes\alpha( b_k)\right)j_{\mathrm{m}}\left(
\sum_{l=1}^na_l\otimes\alpha(b_l)
\right)\Omega:\,n\in{\bf N},\,
\begin{array}{c}a_1,\ldots,a_n\in A\\b_1,\ldots,b_n\in B\end{array}\right\}.
$$ 
\end{lemma}
\begin{proof}(see \cite{II})
Using the Tomita-Takesaki approach one has
\begin{eqnarray*}
\lefteqn{({\bf I}\otimes U_B)\left(\sum_k a_k\otimes b_k\right)j_{\mathrm{m}}
\left(\sum_l a_l\otimes b_l\right)\Omega=}\\
&=&\sum_{kl}a_kj_A(a_l)\Omega_A\otimes U_B b_kJ_Bb_lJ_B\Omega_B\\
&=&\sum_{kl}a_kj_A(a_l)\Omega_A\otimes U_B b_k U_BU_BJ_B
b_l\Omega_B\\
&=&\sum_{kl}a_kj_A(a_l)\Omega_A\otimes U_B b_k U_BJ_BU_B
b_l U_BJ_B\Omega_B\\
&=&\left(\sum_k a_k\otimes \alpha(b_k)\right)j_{\mathrm{m}}\left(
\sum_l a_l\otimes\alpha(b_l)\right)
\end{eqnarray*}
In the third equality we used the fact that $U_B$ commutes with $J_B$.
\qed
\end{proof}

This leads us to:

\begin{tw} (see \cite{II})
In the finite dimensional Hilbert case
$(\jed\otimes U_B){\mathcal{P}}={\mathcal{P}}'$ where
${\mathcal{P}}'$ is the natural cone associated with $(\pi_A(A)\otimes \pi_B(B)',\Omega)$.
\end{tw}
\begin{proof}(see \cite{II})
We just proved, that
$(\jed\otimes U_B){\mathcal{P}}$ is the closure of the set
$$
\left\{\left(\sum_{k=1}^na_k\otimes\alpha(b_k)\right)
j_{\mathrm{m}}\left(\sum_{l=1}^na_l\otimes\alpha(b_l)\right)\Omega:
\,n\in{bf N},\,a_1,\ldots,a_n\in A,\,b_1,\ldots,b_n\in B\right\}.
$$
By Proposition \ref{przestawianie}(\ref{przestawianie2}) 
$\alpha$ maps $\pi_B(B)''$ onto $\pi_B(B)'$, so
the assertion is obvious.
\qed
\end{proof}

Consequently, 
\begin{cor}(see \cite{II})
\label{char2a}
\begin{enumerate}
\item
${\mathcal{P}}_k\cap{\mathcal{P}}_k^\tau$ is nothing else but
${\mathcal{P}}_k\cap{\mathcal{P}}_k^\prime$.
\item
Thus, we got an alternative recipe for constructing a PPT state which is not a separable one.
\end{enumerate}
\end{cor}

\section{Acknowledgments}
The author is greatly indebted to Slava Belavkin and Marco Piani for their valuable remarks.

\end{document}